\documentclass[iop]{emulateapj}
\def\hw{0.49}

\def\fw{0.9}

\usepackage{graphicx, natbib, color, bm, amsmath, epsfig}

\usepackage{graphicx, natbib, color, bm, amsmath, epsfig}

\newcommand{\half}{\frac{1}{2}}

\newcommand{\mach}{\ensuremath{\mathcal{M}}}
\newcommand{\msun}{\ensuremath{\rm{M}_\odot}}

\definecolor{meta}{rgb}{0.371,0.617,0.625} 

\newcommand{\dc}[1]{}

\newcommand{\sci}[1]{\times \ensuremath{{10^{#1}}}}
\def\alf{Alfv\' en}
\def\sa{super-Alfv\' enic}
\def\Sa{Super-Alfv\' enic}

\begin{document}
\title{Mass and Magnetic distributions in Self Gravitating Super Alfv\' enic
Turbulence with AMR.}
\author{David C. Collins \altaffilmark{1}, Paolo Padoan \altaffilmark{2}, Michael
L. Norman \altaffilmark{1}, Hao Xu  \altaffilmark{3}}

\altaffiltext{1}{Center for Astrophysics \& Space Sciences and
Department of Physics, University of California at San Diego, La Jolla, CA}

\altaffiltext{2}{ICREA-ICC, University of Barcelona, Spain }
\altaffiltext{3}{Theoretical division, Los Alamos National Lab, Los Alamos, NM}

\begin{abstract}
In this work, we present the mass and magnetic distributions found in a recent Adaptive Mesh
Refinement (AMR) MHD simulation of supersonic, \sa, self gravitating
turbulence.  Power law tails are found in both mass density and magnetic field
probability density functions, with $P(\rho) \propto \rho^{-1.6}$ and
$P(B)\propto B^{-2.7}$.  A power law relationship is also found between magnetic field
strength and density, with $B\propto \rho^{0.5}$, throughout the collapsing
gas.  The mass distribution of gravitationally bound cores is shown to be in
excellent agreement with recent observation of prestellar cores.  The mass to
flux distribution of cores is also found to be in excellent agreement with recent Zeeman splitting
measurements.

\end{abstract}



%
%
\section{Introduction}\label{sec.introduction}
Understanding the details of star formation is one of the great open problems in
astrophysics today.  One of the central questions is the role of magnetic
fields.  Two opposite paradigms have been explored during the study of this
problem.  The earliest paradigm argues for strong fields that dominate the
process \citep{Shu87,Mouschovias87a,Mouschovias87b}
and yield their dominance by way of ambipolar diffusion; the other extreme
argues for
relatively weak fields that at most alter the thickness of shocks
\citep{Padoan99,Padoan07}, with the majority of the details being determined by the turbulence.


Central to the development of the second paradigm has been the development of
more advance computing hardware and software.  Only a handful of simulations
have been performed that incorporate all three necessary aspects: turbulence, gravity, and
magnetic fields.  Notably are \citet{Gammie03}, \citet{Li04}, \citet{Heitsch01},
\citet{Vazquez-Semadeni05}, \citet{Padoan10}, \citet{Price08,Price10}.  The enormous range of scales at work in this
problem  have motivated the development of high dynamic range techniques,
namely AMR \citep{Fromang06,Collins10} and the addition of MHD to
SPH \citep{Price04}.  These techniques are beginning to yield interesting
results on the nature of MHD in star formation \citep{Price08, Dib10}.

In this work we focus on the distribution of mass and magnetic fields in
\sa\ simulation using high resolution AMR simulations.  Section
\ref{sec.numerics} will introduce the numerical algorithm and discuss the initial
conditions and parameters of the simulation.  In Sections
\ref{sec.densitypdf} and \ref{sec.magneticpdf} we present the volume and mass weighted density PDF for both
density and magnetic field. In Section \ref{sec.fielddensity} we discuss the relation
between mass density and magnetic field strength.  In Section
\ref{sec.mass_to_flux}, we
discuss the relationship between line-of-sight magnetic field strength and
column density (typically used as a proxy for the mass-to-flux ratio $M/\Phi$,
where $M$ is the mass, and $\Phi$ is the magnetic flux), and compare to recent observations.  In Section
\ref{sec.massdist}
we present the mass distribution of cores, and compare to observations of
protostellar cores.  In Section \ref{sec.conclusions} we will conclude.

\section{Numerical Model}\label{sec.numerics}

\subsection{Simulation software}
The simulation was performed with the MHD extension of Enzo by
\citet{Collins10}.  It employs the MHD Godunov solver of \cite{Li08a} to solve the MHD 
equations; to constrain $\nabla\cdot B$ it employs the CT scheme of
\cite{Gardiner05}; the AMR refinement is done with the divergence
free interpolation of \citet{Balsara01}.  An isothermal equation of state and
ideal MHD are assumed throughout.

Refinement is performed such that the Jeans length,
$$
\lambda_J =\sqrt{ \frac{\pi c_s }{G \rho}} ,
$$
where $c_s$ is the sound speed and $G$ is the gravitational constant, is
resolved by four grid zones
everywhere.  Thus, whenever the density in a zone is above the "Truelove Density"
$$\rho_T = \frac{\pi^2 c_s^2 }{16 G \Delta x^2}, $$
that zone is flagged for refinement.  The maximum number of levels is 4. With a base
grid of $128^3$, this gives a maximum effective resolution of $2048^3$.

\subsection{Simulation Parameters}\label{sec.parameters}
We began with a uniform density and magnetic field with no self gravity, and
stirred with a Gaussian random field with no compressional modes, with power
distributed in a top-hat between between $1 \le k \le
2$ and mean Mach number 8.9.   Driving is done in the same manner as
\citet{MacLow99}, such that the energy injection rate is constant.  After a number of dynamical times, gravity was activated.
This time is $t=0$.  Driving continued for the duration of the collapse.

The initial uniform fields have a ratio of gas to magnetic pressure of $\beta$
=$8 \pi c^2 \langle \rho \rangle / \langle B^2 \rangle = 22.2$, but the mean
squared field
strength has been amplified by the driving so that when gravity is turned on at
$t=0$, $\beta$ = 0.2.  This yields an Alfv\' en Mach number of 2.8 at $t=0$.
have been chosen so that $\alpha = 5 \sigma^2 R/(3 G M)
\approx 1$, where $\sigma$ is the three dimensional velocity dispersion, $M$ is the total mass in
the box, and $R$ is half the width of the box. The typical parameter used to measure the relative effects of
magnetic and gravitational energy is the mass-to-flux ratio in units of the
critical value for support.  If we use the critical value for a sheet, as in
\citet{Nakano78}, $(M/\Phi)_c = (4 \pi^2 G)^{-1/2}$, the uniform
(pre-turbulence) cube gives us
\begin{align}
\lambda = (M/\Phi)/(M/\Phi)_c = 18.7. 
\end{align}

Mass to flux is a reasonable quantity to parameterize magnetic support for simple
geometries, like spheres and disks.  However, for situations like the one
presented here, in both the turbulent box at $t=0$ and the prestellar cores
discussed later, the spatial structure of the objects in question are not
easily defined.   We find it more instructive to examine the ratio of energies.  For sake
of comparison to $M/\Phi$, we define
\begin{align}
\lambda_E = \sqrt{E_G/E_B},  
\end{align}
where
\begin{align}
E_G & = G \int \int \frac{\rho(r) \rho(r^\prime)}{|r-r^\prime|} d^3 r d^3 r^\prime
\\
E_B &= \int \frac{B^2}{8 \pi} d^3 r.
\end{align}
For the initial conditions at $t=0$ (after the initial turbulent evolution, but
before gravitational collapse), we have 
\begin{align}
\lambda_E = 6.5
\end{align}
For reference, a sphere with critical mass $M_c = \Phi /(2 \pi \sqrt{G})$,
$\lambda_E=0.2$.  

In a similar vein, if we define $\alpha_E = 2 E_K/E_G$ as the
ratio of measured gravitational energy to total kinetic energy $E_K =\half \int \rho
v^2 d^3 r$, we find for our data at $t=0$
\begin{align}
\alpha_E =0.64.
\end{align}


\subsection{Physical Scaling}\label{sec.units}
Although these simulations are scale free, we use a box
size of 10 pc and density of $300 \mathrm{cm}^{-3}$ for the analysis presented
here.   This gives a resolution on the finest zones of 1000 AU and a mean mass
density $\rho_0=1.2\sci{-21}$ g.  The sound
speed is set to be 0.2 $\mathrm{km s}^{-1}$, giving a temperature of $\approx
10$ K.  Mean and RMS magnetic fields are 0.6 and 2.7 $\mu G$, respectively, with
the mean along the $\hat{z}$ direction.  These units give a total mass of
$1.2\sci{4} \msun$

Rescaling can be found with the
following relations:
\begin{align}
t_{\rm{ff}} = &~ \sqrt{ \frac{3 \pi}{32 G \rho}} \label{eqn.freefall_constraint}\\
L \propto &~ T^{1/2} \rho^{-1/2} \label{eqn.length_constraing}\\
M \propto &~ T^{3/2} \rho^{-1/2} \label{eqn.mass_constraint}\\
c_s \propto &~ T^{1/2}  \label{eqn.soundspeed_constraint}\\
B \propto &~ T^{1/2} \rho^{1/2}.
\end{align}

\subsection{Core definition}\label{sec.cores}
We select cores with the algorithm
described by \citet{Smith09} which selects topologically connected density isosurfaces at a
spacing of $\delta \rho = \rho_{i}/\rho_{i-1}$ that have no substructure; that
is, a contour at $\rho_{i-1}$ that contains zero or one contour above $\rho_{i}$.  This is a
similar definition to that used by \citet{Padoan07} and \citet{Schmidt10}, both of whom
reported no significant effects due to variations in $\delta \rho$, provided
$\delta \rho < 1.16$. In our
simulation, $\delta \rho = 1.12$.  A core is
determined to be bound if its gravitational energy is at least twice as large as the sum
of kinetic, thermal, and magnetic energies. All analysis has been performed with
the AMR analysis package {\tt yt} \citep{Turk08}.
 
\begin{figure*} \begin{center}
\includegraphics[width=\hw\textwidth]{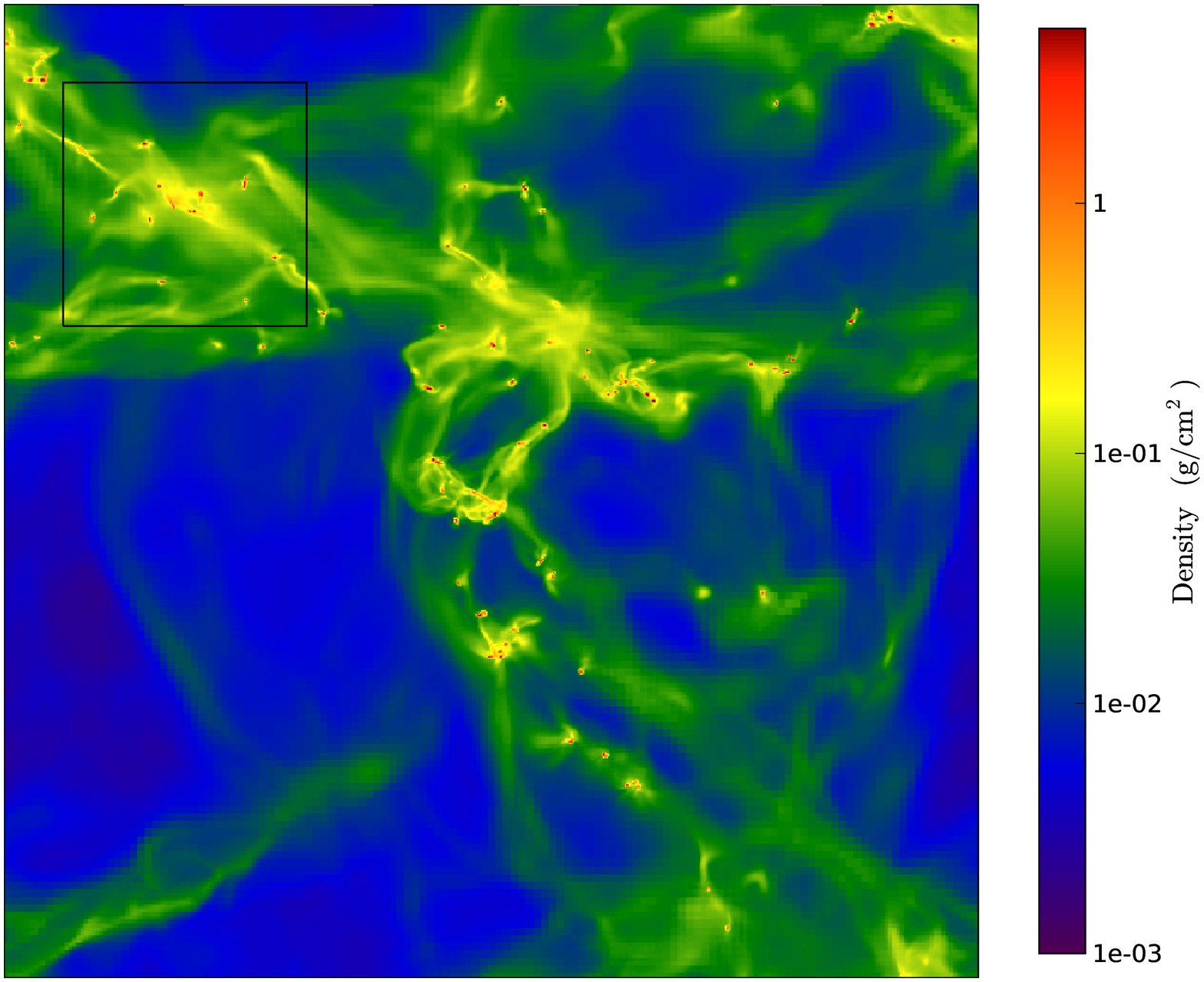}
\includegraphics[width=\hw\textwidth]{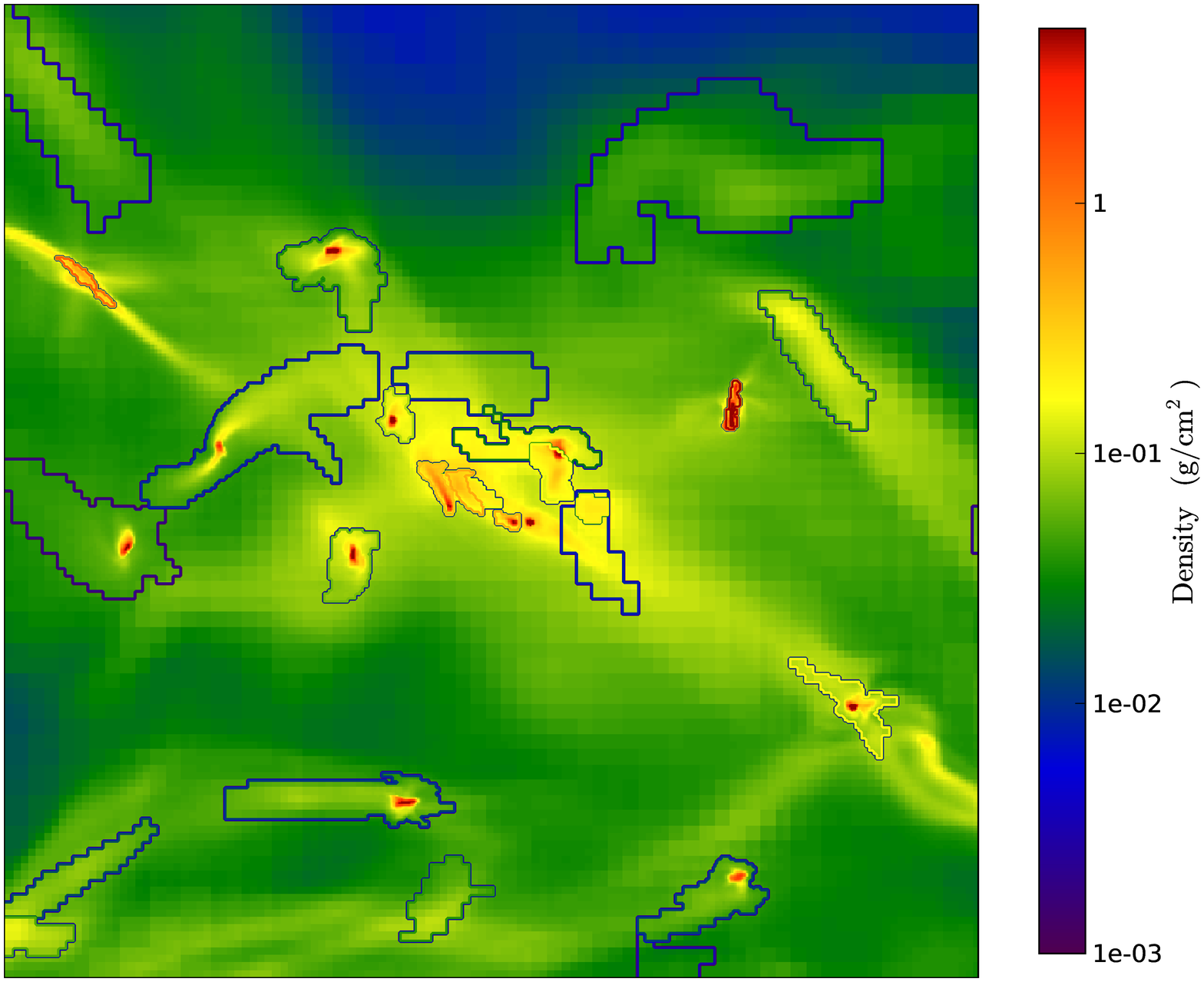}
\caption[ ]{Projection of the full domain at $t=0.75 t_{ff}$ (left) and a
close up of the squared region, including curves indicating the edges of all
bound cores in the region (right).}
\label{fig.projection} \end{center} \end{figure*}

All analysis in this section takes place at a single snapshot, $t=0.75 t_{ff}$.
At this time, there are 148 cores that match the definition of bound. Figure
\ref{fig.projection} shows a projection of density at this timestep in the left
panel, and a close up of a portion of the column density map in the right panel
with core boundaries drawn.
All cores drawn are associated with topologically isolated, gravitationally
bound objects, but some have their features washed out by projection effects.

\section{Density PDF}\label{sec.densitypdf}
One of the most prominent consequences of supersonic turbulence is the log-normal distribution of densities \citep{Vazquez-Semadeni94,
Padoan97a, Padoan97b,
Scalo98,Passot98,Nordlund99,Klessen00,Padoan02,Padoan10,Federrath08}.
This has been used to predict several properties of star formation, including the Initial Mass Function of stars
(IMF) \citep{Padoan02,Padoan07}, brown dwarf frequency \citep{Padoan04} and the star formation rate
\citep{Krumholz05}.  Here we will discuss the PDFs one expects to see
from isothermal turbulence, and what has been seen in our simulations with the
inclusion of self gravity.

The central limit theorem states that the sum of a sufficiently large
number of uncorrelated events will form a Normal, or Gaussian,
distribution.  A corollary of this is a sufficiently large number of
random {\it multiplicative} events will form a lognormal distribution.  This
distribution has been experimentally verified in a large number of different simulations,
both pure hydro \citep{Vazquez-Semadeni94,Padoan97b, Kritsuk07, Kritsuk09b} and MHD
\citep{Li04,Lemaster08, Kritsuk09b}

The log-normal distribution is given by 
\begin{align}
P(x) d\ln x = \frac{1}{\sqrt{2 \pi \sigma^2} }~ exp\left[ \frac{(\ln x -
    \mu)^2}{2 \sigma^2} \right ] d\ln x
\label{eqn.lognormal}
\end{align}
where $x = \rho/\rho_0$ is the over density, and $\mu =
-\sigma^2/2$ is the mean of $\ln x$.  For pure hydrodynamical
turbulence, 
\begin{align}
\sigma = \sqrt{\ln(1+b^2 \mathcal{M}^2)}
\label{eqn.lognormalWidth}
\end{align}
where $b$ has been determined numerically to lie between 0.3 and 0.4
\citep{Padoan97b,Federrath08, Kritsuk07, Beetz08, Kritsuk09a, Federrath10}. The value $b$ has been
shown to vary by as much as a factor of 3, depending on the ratio of solenoidal
to compressive forcing \citep{Federrath08, Federrath10}.

For driven
MHD turbulence, \citet{Lemaster08} find that
\begin{align}
\sigma_{\rm{LM08}} = \sqrt{|-0.72 \ln\left[1+0.5 \mathcal{M}^2\right] + 0.20|},
\end{align}
and is insensitive to the magnetic field strength.

\begin{figure*}
\begin{center}
\includegraphics[angle=0,width=\hw\textwidth]{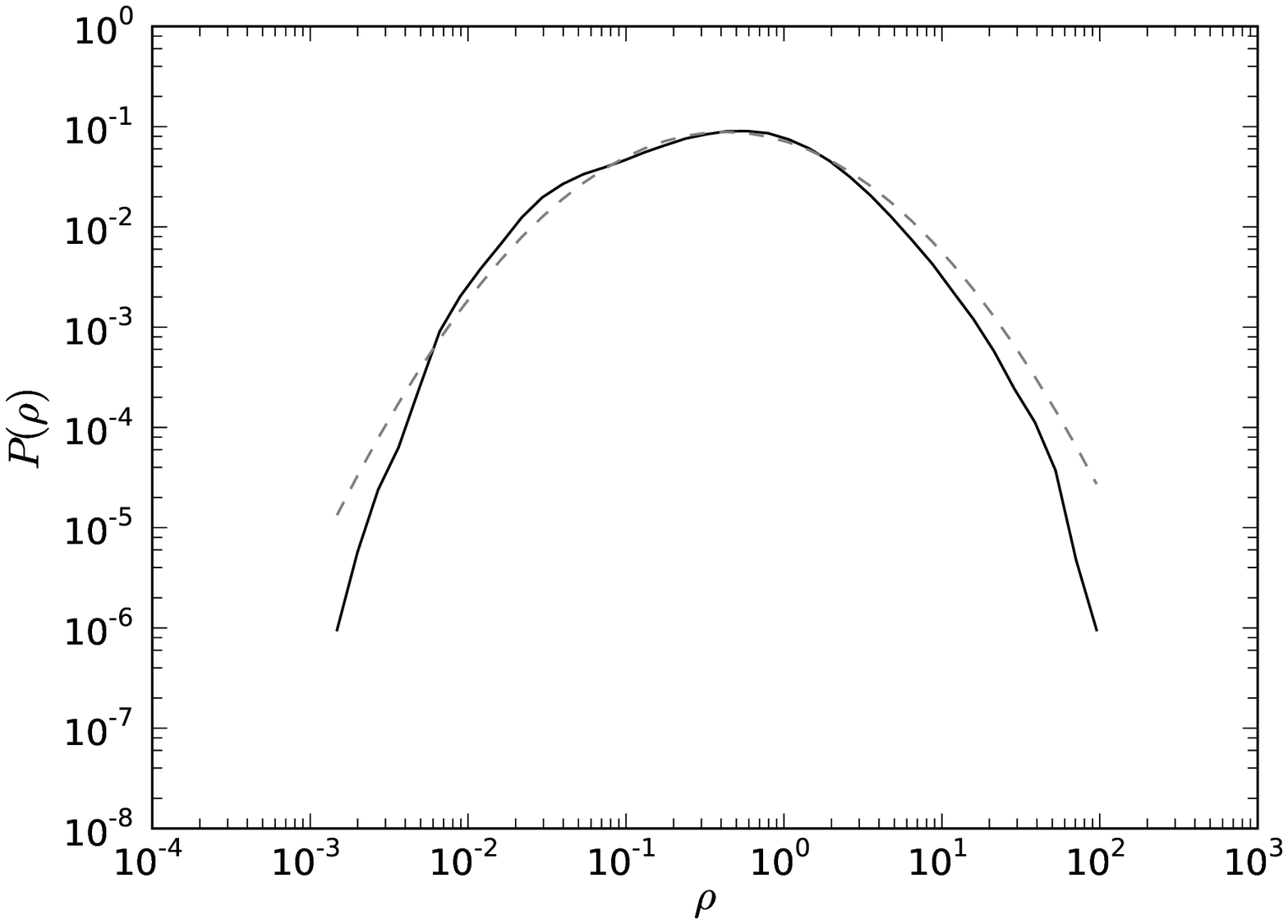}
\includegraphics[angle=0,width=\hw\textwidth]{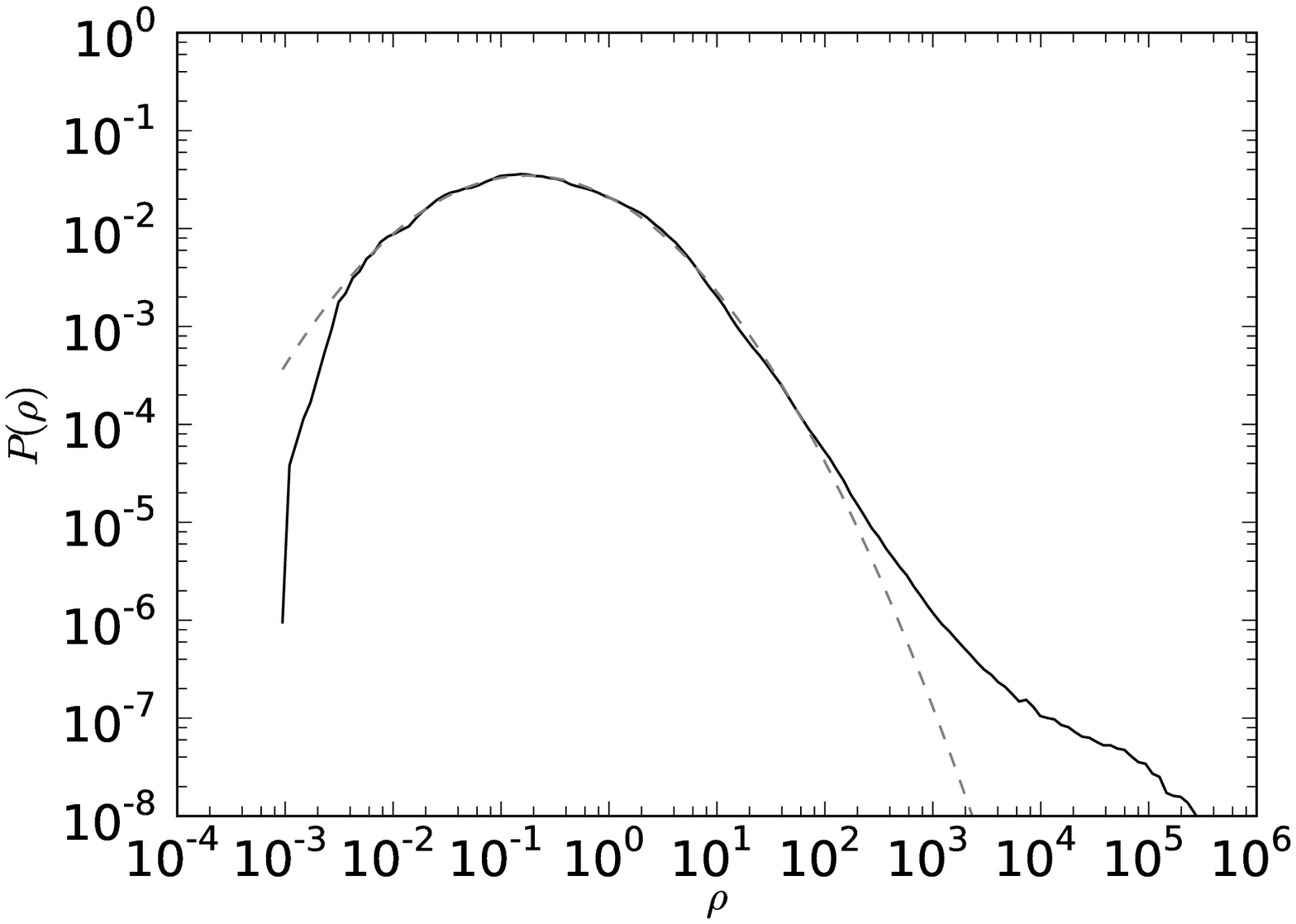}
\caption[Density PDFs with Log Normal fits]{Density PDFs of the
  initial conditions (left) and after $t=0.75 t_{ff}$ (right). Both are fit to
  lognormal distributions (dashed lines).  }
\label{fig.LogNormalFits}
\end{center}
\end{figure*}

Figure \ref{fig.LogNormalFits} shows density PDF $P(\rho)$ for two snapshots:
at $t=0$ (left) and $t=0.75 t_{ff}$ (right). In both plots, the solid line is
the measured PDF and the dashed line is the fit to a lognormal.  Table  \ref{table.ok4LogNormalFits} shows the fit
parameters.  The addition of self gravity causes the PDF to widen and the mean
to decrease.   We find that $b$ = 0.3 for our initial
conditions, similar to \citet{Kritsuk07}, but $b$ = 0.5 for the collapsed snapshot; we also find that the
dispersion, $\sigma_{\rm{LM08}}$, found  by \citet{Lemaster08} is in better agreement with
the collapsed state of the simulation.

When turbulence simulations are performed in the presence of self
gravity, several authors \citep{Klessen00,Slyz05,Vazquez08, Federrath08b,Kritsuk10} find that
the log-normal PDF underestimates the high density tail of the measured PDF.  \citet{Slyz05} fit the high density tail to a power law with index of $-1.5$.
\citet{Klessen00} does not fit a power law, but  
the resolution of those simulations is much lower than what we present here.
\citet{Vazquez08} mention the existence of a power law, but say nothing further.
Observationally, \citet{Kainulainen09} found power law
wings in column density distributions of active star forming regions, and a
similar high density power law has been seen in Aquila (Philippe Andr\' e,
private communication.), Our work is the first reported case in an MHD simulation.

\citet{Kritsuk10} find an extended two part power law, with index $-1.7$ at intermediate
densities, and $-1$ at high densities.  They provide the first explanation of
this power law, associating it with a self-similar singular isothermal sphere.
In such a sphere, $\rho \propto r^{-2}$, thus $V(\rho)\propto \rho^{-3/2}$.

\begin{figure*} \begin{center}
\includegraphics[width=\hw\textwidth]{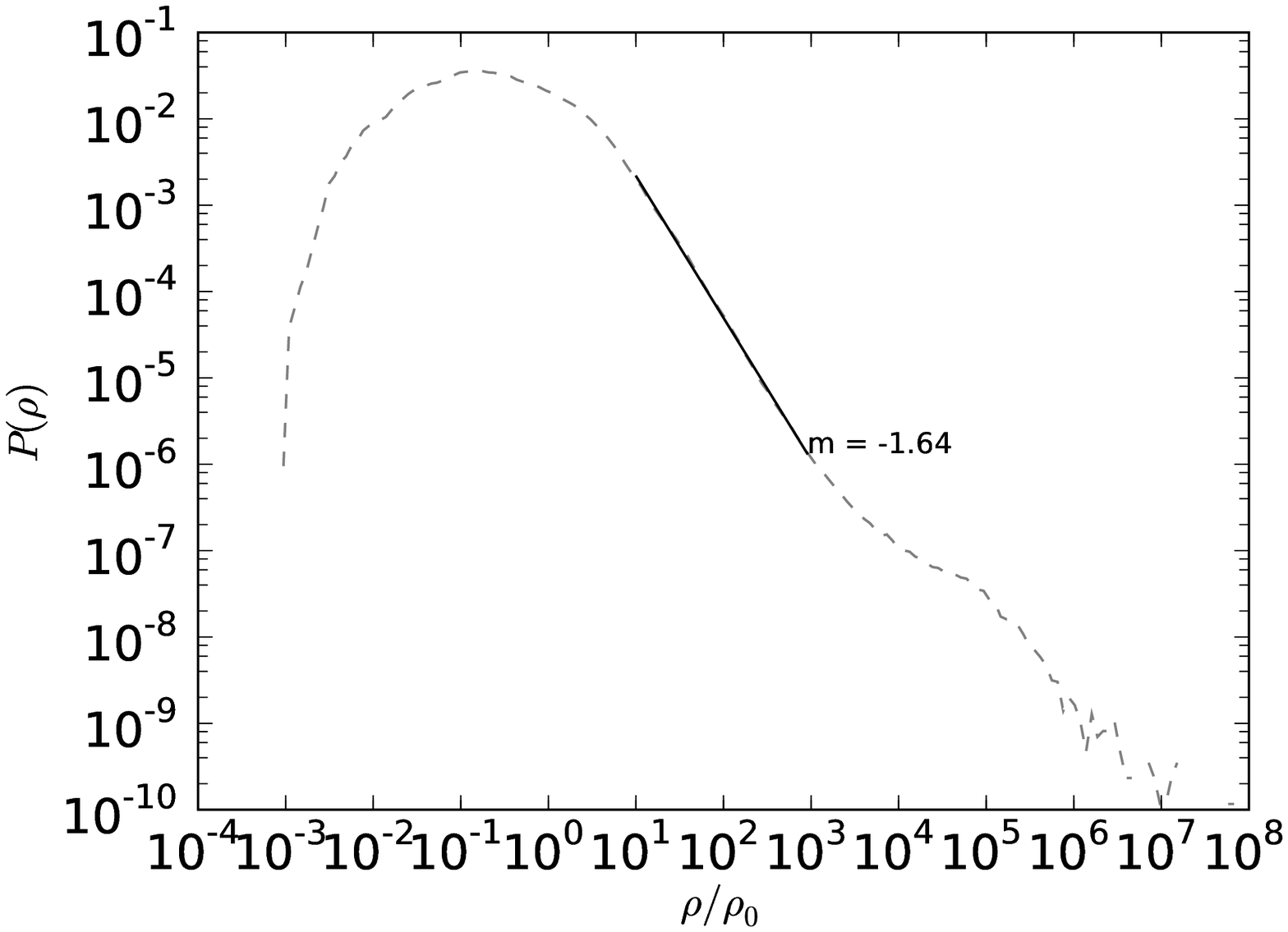}
\includegraphics[width=\hw\textwidth]{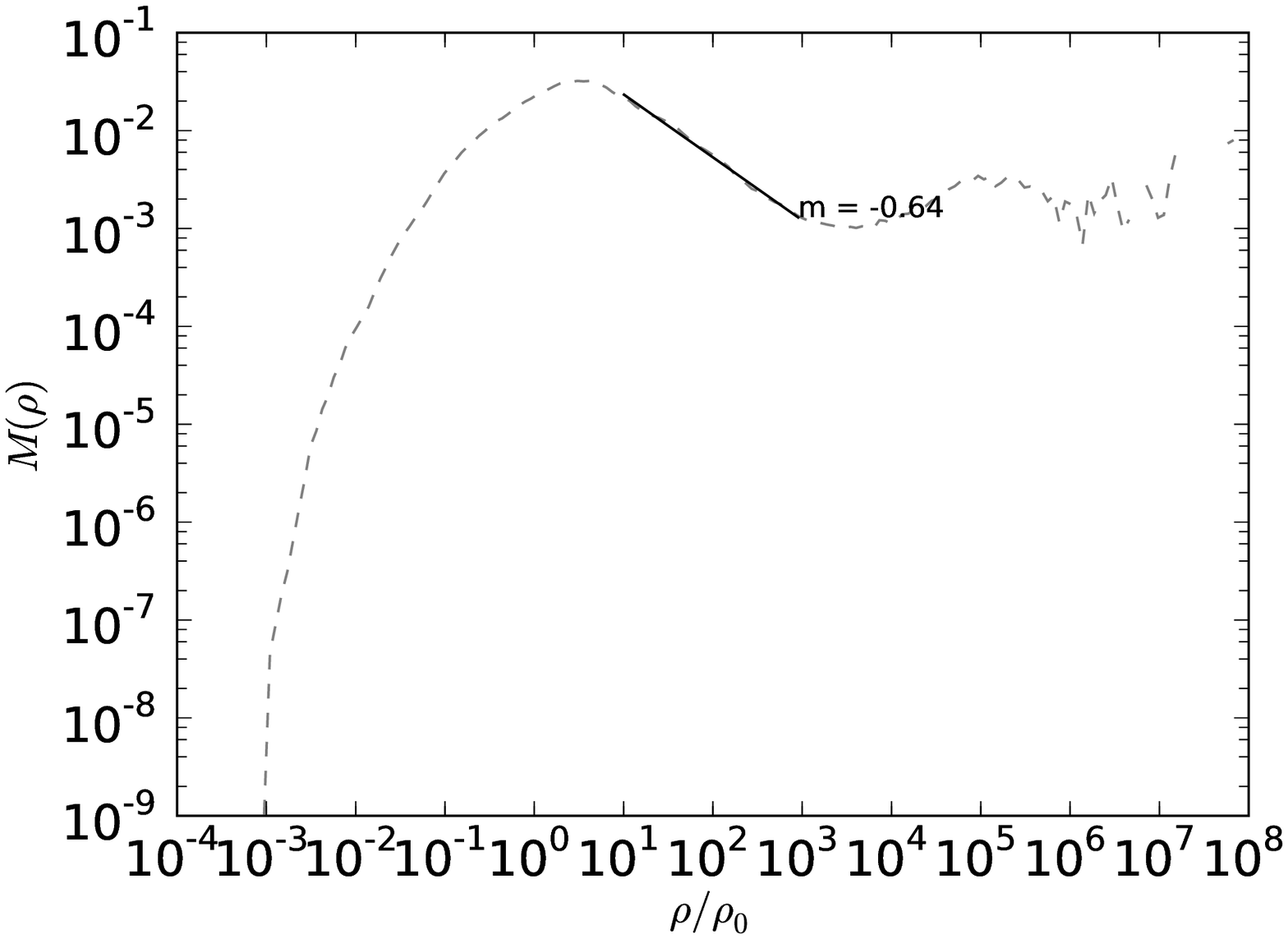}
\caption[ ]{Volume (left) and mass (right) weighted density PDFs at $t=0.75
t_{ff}$, with power law fits.}
\label{fig.PowerLaw} \end{center} \end{figure*}

Figure \ref{fig.PowerLaw} shows both volume-weighted PDF $P(\rho)$ and
mass-weighted PDF $M(\rho)$, with the
power law fits of $-1.64$ and $-0.64$, respectively, in the range of
$\rho=10-1000$.  The dashed line in Figure \ref{fig.PowerLaw} is the same curve
as the solid line in Figure \ref{fig.LogNormalFits}, but here with the mid to high density
power law emphasized.  This power law breaks down above a density of 1000,
likely due to resolution effects, as the maximum resolvable density in the
simulation, according to the Truelove condition \citep{Truelove97}, is a
$\rho/\rho_0=1623$.  The exponent we find is quite close to the $-1.7$ found
by \citet{Kritsuk10}, and in reasonable agreement with the $-1.5$ they predict
from a singular isothermal sphere.   \citet{Federrath08b} also measure the
density PDF from a driven, self-gravitating turbulence simulation (without
magnetic fields) with a set of
Lagrangian tracer particles.  The tracer particles in their work also follow a
power law tail, with an index of $-0.6\pm0.1$, (Fedderath 2010, private
communication) consistent with our mass-weighted
$M(\rho)$.  

\begin{table}
\begin{center}
\caption[Fit parameters for log normal fits for lognormals shown in \ref{fig.LogNormalFits}]{Fit parameters.}
\begin{tabular}{l l l l l l l}
\label{table.ok4LogNormalFits}
$t/t_{ff}$ & \mach & $\mu$ & $\sigma$ & $\chi^2$ & $b$ & $\sigma_{\rm{LM08}}$ \\
\hline
0    & 8.15 &-0.80 & 1.35 & $4.6\times 10^{-6}$ & 0.27 & 1.67\\ 
0.75 & 8.57 & -1.86 & 1.74 & $1.6 \times 10^{-6}$ & 0.52 & 1.65  
\end{tabular}
\end{center}
\end{table}

\section{Magnetic PDF}\label{sec.magneticpdf}

\begin{figure} \begin{center}
\includegraphics[width=\hw\textwidth]{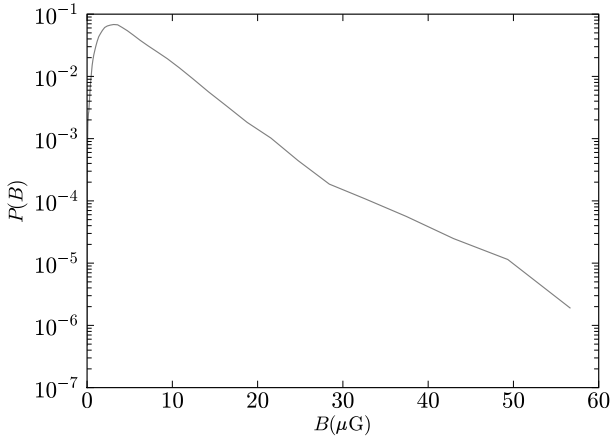}
\caption[ ]{Semi-log plot of $P(B)$ at $t=0$.  A roughly exponential tail can be
seen at high field strength, consistent with earlier \sa\ turbulence simulations.}
\label{fig.Bdist0linlog} \end{center} \end{figure}

Figure \ref{fig.Bdist0linlog} shows the volume-weighted PDF of the magnetic field at
$t=0$ in a semi-log plot.
In agreement with \citet{Padoan99}, we find a roughly exponential tail at high
field strength.  This intermittent distribution in the magnetic field is caused by the large field
amplifications by the strong, three dimensional compressions made possible by the large kinetic
energy, relative to the magnetic energy, of the \sa\ turbulence.

\begin{figure*} \begin{center}
\includegraphics[width=\hw\textwidth]{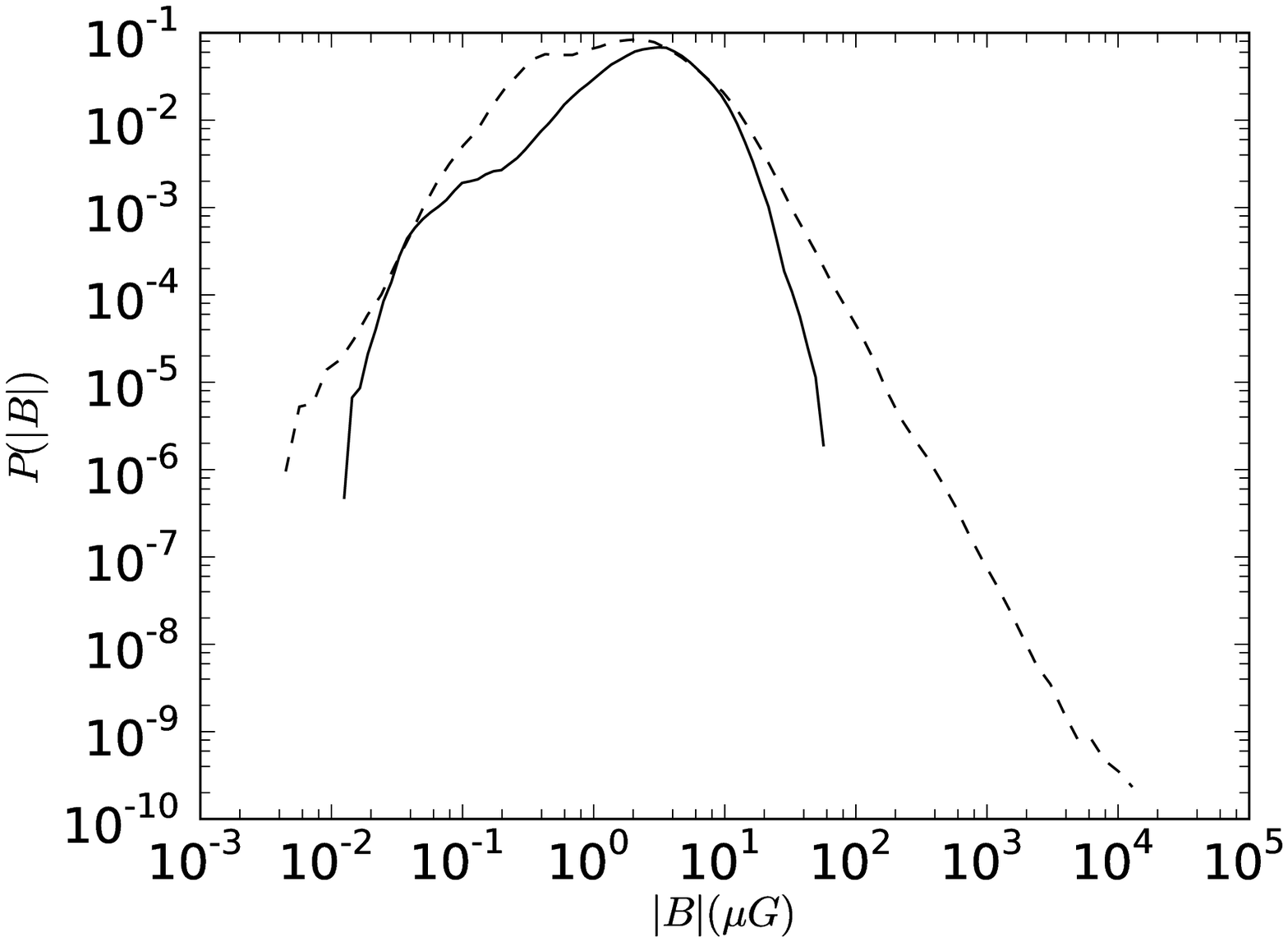}
\includegraphics[width=\hw\textwidth]{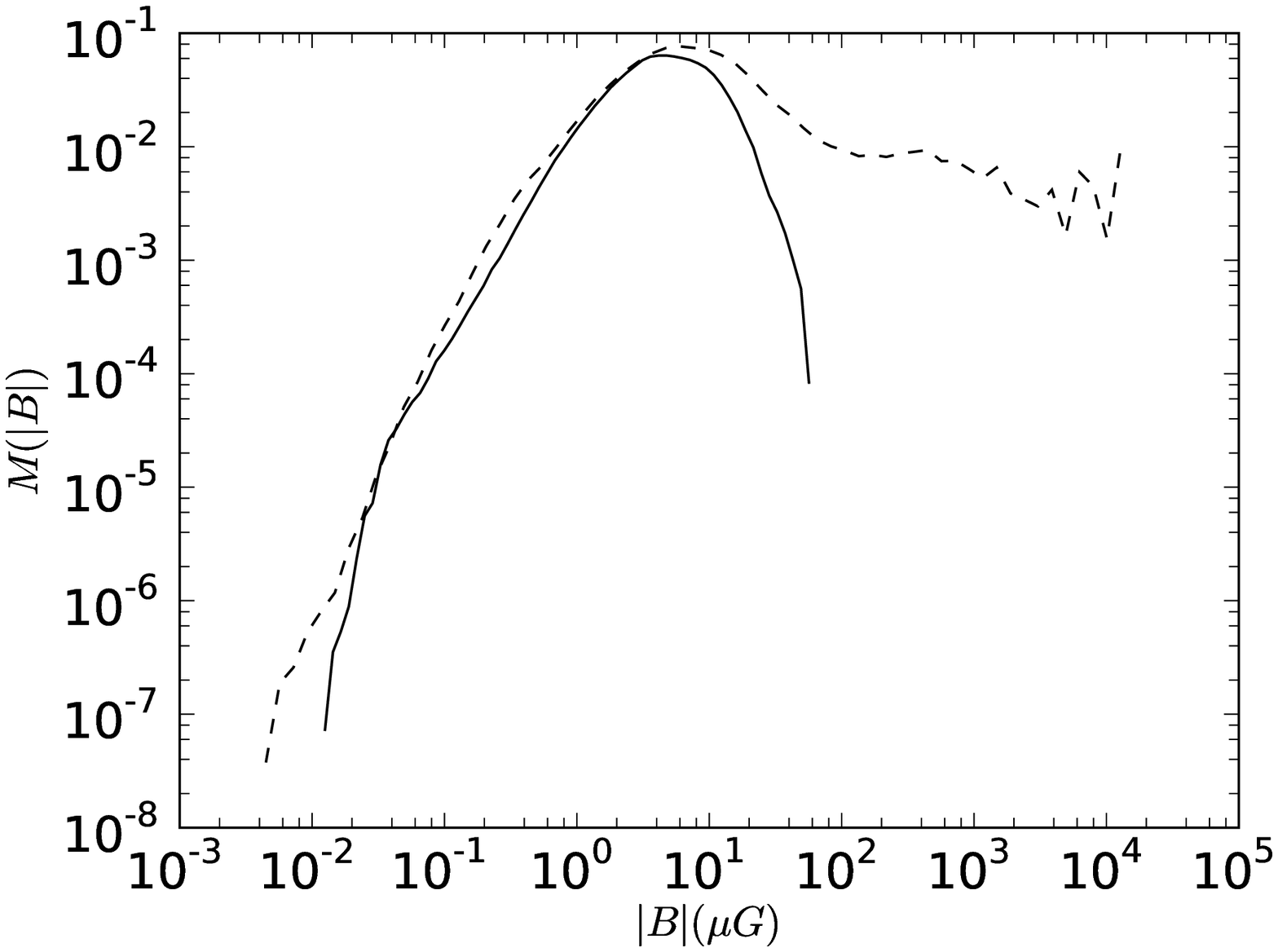}
\caption[ ]{Magnetic field PDF $P(|B|)$ (left) and mass-weighted PDF $M(|B|)$
(right) for $t=0$ (solid line) and $t=0.75 t_{ff}$
(dashed line).}
\label{fig.MagneticPDF} \end{center} \end{figure*}

The left panel of Figure \ref{fig.MagneticPDF} shows the probability density function for the
magnetic field strength, $P(B)$ for $t=0$ (solid line) and $t=0.75 t_{ff}$
(dashed line), here in a
log-log plot.  The most significant aspect of this figure
is the strong power law tail after the collapse has evolved.  Gravitational
collapse amplifies the peak magnetic field strength by 2 orders of magnitude (in
fact a factor of 320),
and creates a prominent power law tail.  This powerlaw tail is fit by
$P(B)\propto B^x$, with $x=-2.74$.   If one naively takes $P(\rho) \propto
\rho^{-1.5}$ as expected from a singular isothermal sphere (as in Section
\ref{sec.densitypdf}) and $\rho \propto
B^2$ (as in other simulations \citep{Li04} and observations
\citep{Bertoldi92, Crutcher99} of dense cores), one arrives at $x=-3$, which is
quite close to the value we find here.  Details will be discussed in the next
Section, where we measure the relationship between $B$ and $\rho$ in our
simulation.

The right panel of Figure \ref{fig.MagneticPDF} shows the mass-weighted PDF,
$M(B)$ for the same snapshots as the left panel.  If we fit a power law to the
same field strength range that was used for $P(B)$, we find $M(B)\propto B^m$,
where $m=-0.4$.  However, the power law in $M(B)$ is not nearly as well defined
as for $P(B)$.  This is due two related effects: the power law relation between
$\rho$ and $B$ is less well defined at field strengths above $B=100 \mu\rm{G}$ (see
Figure \ref{fig.BrhoFit} in the next section); and the power law in density
breaks down above $\rho/\rho_0 > 1000$, likely due to resolution effects (see
Figure \ref{fig.PowerLaw}).  

\section{Field Strength vs. Density}\label{sec.fielddensity}

\begin{figure*} \begin{center}
\includegraphics[width=\fw\textwidth]{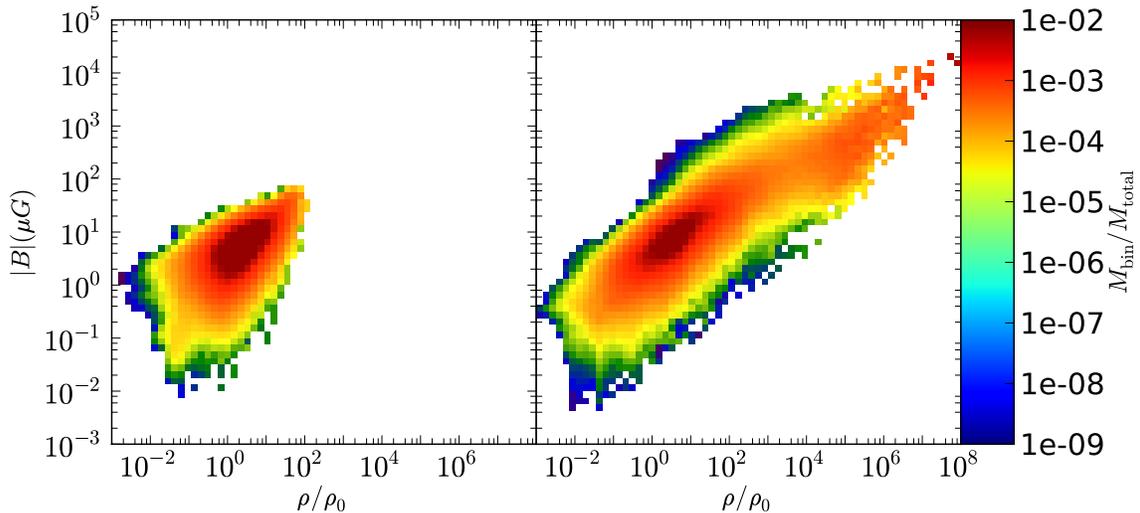}
\caption[ ]{Magnetic field strength vs. density for all zones in the simulation
for $t=0$ (left) and $t=0.75 t_{ff}$ (right).
Color field shows total mass fraction in each $(B,\rho)$ bin.}
\label{fig.Brho} \end{center} \end{figure*}

Figure \ref{fig.Brho} shows a contour plot of magnetic field strength vs
density, colored by fraction of mass in each $(B,\rho)$ bin.  The left panel
shows $t=0$, before the action of gravity.  As in \citet{Padoan99}, the upper
envelope is matched by a power law roughly of the form $B\propto \rho^{0.4}$, and
the scatter is quite large in both $B$ and $\rho$.  The large scatter in field
strength is due to the fact that only the component of the field perpendicular
to a shock is amplified, but due to the weak nature of the magnetic field, the
relative orientations of $\vec{B}$ and the shock are not correlated.  \citet{Padoan99}
demonstrated that in models with larger magnetic field strength (sub-Alfv\'
enic turbulence), the flow is constrained to move primarily along the field
lines, significantly decreasing the field strength amplification power of the
turbulence, and decreasing the scatter in the magnetic field strength. The right panel
in Figure \ref{fig.Brho} shows the $B-\rho$ relation at $t=0.75 t_{ff}$, after gravity has taken effect.
Several features are noticeable.  Both $B$ and $\rho$ are amplified by almost
four orders of magnitude by the gravity, and the initial scatter in the
distribution remains imprinted on the self gravitating distribution.  The
relation gets more shallow at $\rho/\rho_0 > 10^3$, which is the same density
at which the power law slope seen in $P(\rho)$ changes.  As before, this is likely due to
resolution effects.  Another feature of figure \ref{fig.Brho} is that the upper envelope
slope increase sharply at $\rho/\rho_0 > 5$.  Above this density, gravitational
collapse has overtaken the dynamics, as evidenced by the onset of the power law
in the density PDF.  As these motions are three dimensional contractions, rather
than the one dimensional compression due to shocks, the amplification is
stronger than what is possible from the turbulence alone.


\begin{figure} \begin{center}
\includegraphics[width=\hw\textwidth]{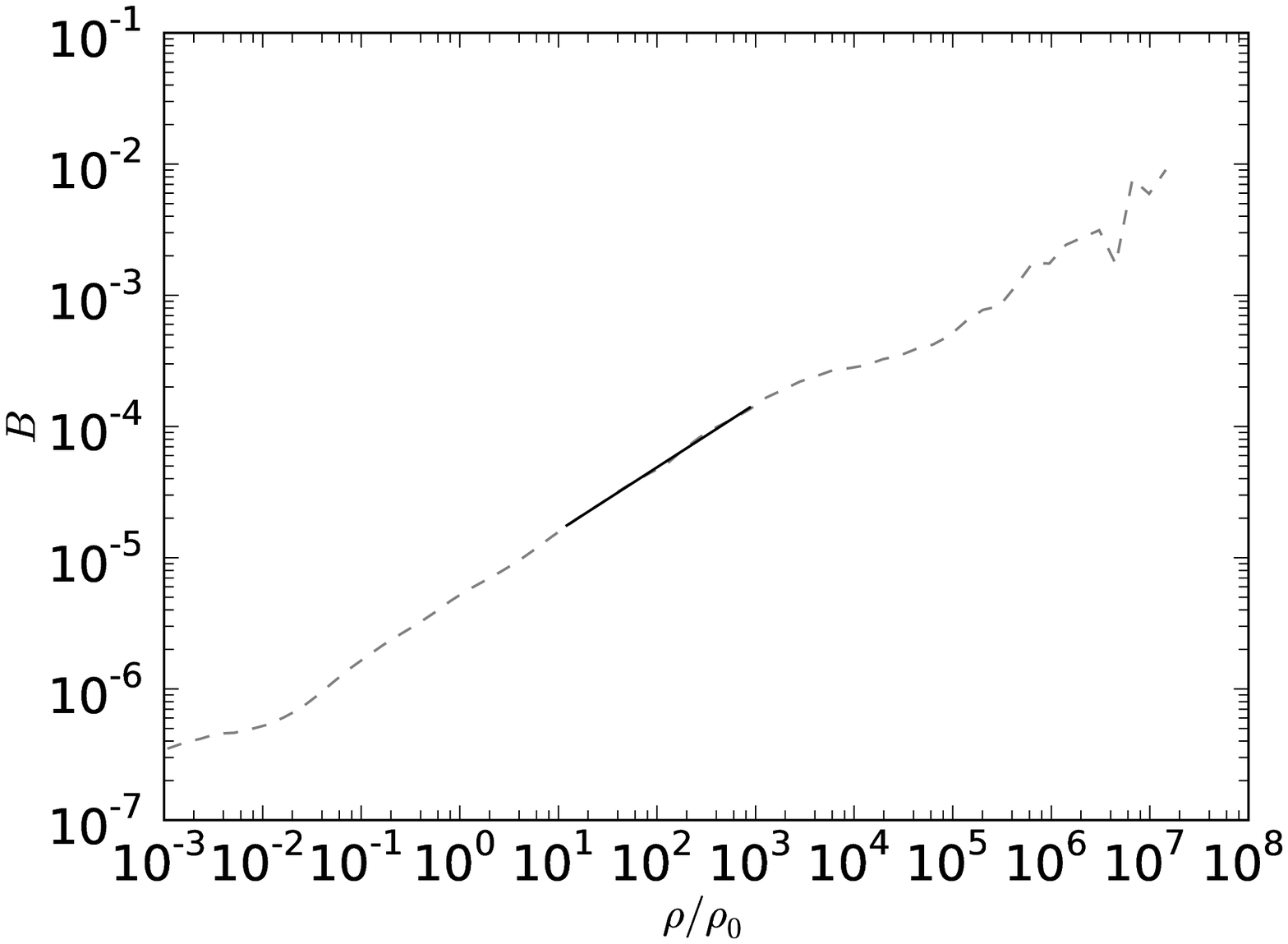}
\caption[ ]{Mass weighted average $B$ vs. $\rho$ (dotted line) and fit to the
range where the density PDF is also a clear power law (solid
line).  The fit is a power law with exponent 0.48. }
\label{fig.BrhoFit} \end{center} \end{figure}

Figure \ref{fig.BrhoFit} shows a mass weighted average $B$ as a function of
density bin, which has been fit to  a power law
between over densities of 10 and 1000.  We find $B\propto \rho^{0.48}$. This behavior has
been predicted or observed by several other authors: \citet{Fiedler93} predicted
$B \propto \rho^\kappa$, with $\kappa=0.44-0.5$, though this was done in a
quasi-static collapse model with ambipolar diffusion; both \citet{Bertoldi92} and
\citet{Crutcher99} found a similar result in dense molecular cores and interpreted it as constancy of the
\alf\ speed; \citet{Li04} found this behavior for central density and magnetic
field strengths.  Figure \ref{fig.Brho} demonstrates that this behavior is
endemic to the entire collapse process, not just the high density collapsed
objects.  

\section{Mass to flux ratio}\label{sec.mass_to_flux}

\begin{figure*} \begin{center}
\includegraphics[width=\fw\textwidth]{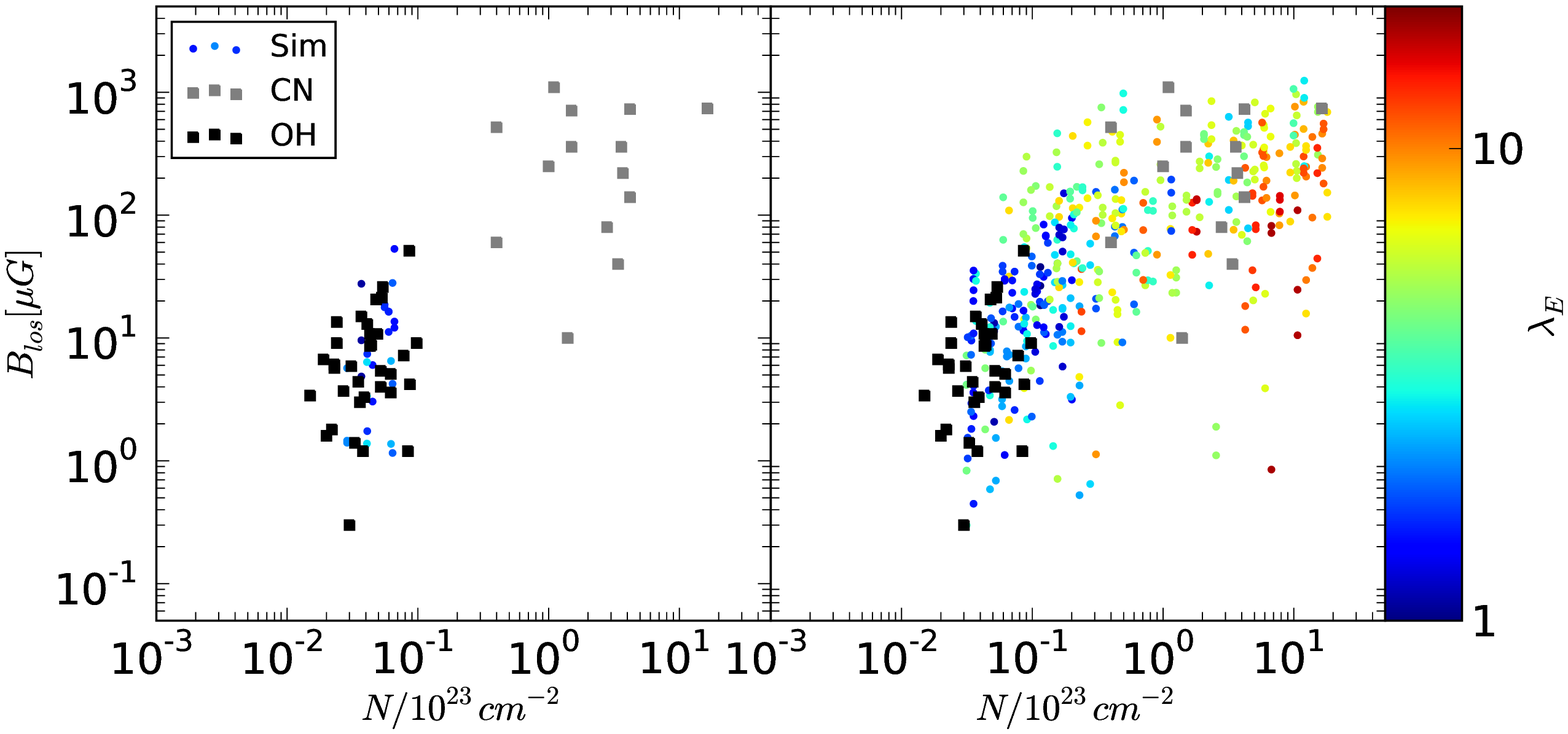}
\caption[ ]{Line of sight magnetic field strength $B_{\rm{los}}$ vs column density
$N$ for three populations of cores: CN Zeeman splitting measurements of
\citet{Falgarone08} (grey squares); OH Zeeman splitting measurements of
\citet{Troland08} (black squares); and
bound objects in our simulation (colored points).  Color indicates the ratio of gravitational to
magnetic energy. At time $t=0$ (left) and $t=0.75 t_{ff}$ (right)   Color indicates $\lambda_E = \sqrt{E_G/E_B}$}\label{fig.BlosN} \end{center} \end{figure*}

Figure \ref{fig.BlosN} shows line-of-sight magnetic field strength, $B_{\rm{los}}$,
vs column density,
$N$, for three populations of cores: CN Zeeman splitting measurements of
\citet{Falgarone08}; OH Zeeman splitting measurements of \citet{Troland08}; and
bound objects in our simulation.  Color shows $\lambda_E=\sqrt{E_G/E_B}$, our
adopted proxy for mass to flux ratio.  
 The left plot shows $t=0$, the right plot shows $t=0.75 t_{ff}$.  In the simulated points, the line of sight is taken along each
of the three coordinate axes, so that there are three points in the plot for each simulated
core; $B_{\rm{los}}$ is the density weighted average of each
field quantity, and $N=M/A$, the total computed mass of the core divided by the
area projected along that axis.  The observational points are currently the best data available
to relate magnetic field strength to mass, and some of the only magnetic field
measurements for high density protostellar cores.  The ratio between
these quantities is often used as a proxy for mass-to-flux ratio $\lambda$
as 
\begin{eqnarray}
\lambda = c_g \frac{N(H_2)}{B_{\rm{los}}} \frac{\sqrt{G}}{c_\Phi},\label{eqn.Bourke01}
\end{eqnarray}
where $c_g = (1/2, 1/3)$ is a geometrical correction for (spherical, sheet)
projection effects, and $c_\Phi = (0.12, 1/2\pi)$ is a correction found numerically for
equilibrium configurations of a (sphere, sheet), respectively \citep{Bourke01}.
Here, we directly compare $B_{\rm{los}}$ and $N(H_2)$. This allows us to eliminate
the need for either correction factor and compare the results of our model
directly to the observations.

We find that our cores and the
observed cores have almost the same distribution in the $B_{\rm{los}}$-$N$ space.
This shows that the early evolution of prestellar cores is well reproduced with
isothermal super Alfv\' enic turbulence
and self-gravity.  Simulations without self gravity by \citet{Lunttila08}
successfully reproduce the lower density OH measurements, but lack the density
range to reproduce the CN observations.  This is also seen in the left
panel of Figure \ref{fig.BlosN}, which shows the initial time $t=0$, at which only the
effects of turbulence are felt by the gas.  The inclusion of self gravity and
the large range of density scales allowed by AMR reproduces the higher density
CN points.  

The color of the simulated points corresponds to $\lambda_E=\sqrt{E_G/E_B}$,
non-spherical analog of $\lambda=(M/\Phi)/(M/\Phi)_c$.  \citet{Li04} found that
$\lambda > 10$ for all cores in question.  We find that $\lambda_E > 1$ for all objects,
which, using the spherical case as a guide, is analogous to $\lambda> 5$, so
our results are in reasonable agreement with theirs.



The simulated points at $t=0.75 t_{ff}$ in figure \ref{fig.BlosN} are best fit by a
power law
\begin{eqnarray}
B_{\rm{los}} = & N^{0.57} \label{eqn.BN}
\end{eqnarray}
Collapse that preserves mass-to-flux would have an exponent of 1, by equation
\ref{eqn.Bourke01}.  Since ideal MHD preserves $M/\Phi$ along a flux tube, this
indicates that flow along the field lines must be responsible for some of the
dynamics.  As discussed by \citet{Padoan99}, this is due predominantly to flow
along the magnetic field lines.  This flow is due predominantly to kinematic
alignment between the velocity and magnetic field, wherein the kinetic energy
stretches the magnetic field, hence aligning the two.  The relationship in
equation \ref{eqn.BN} is also expected from figure \ref{fig.Brho}, 
demonstrating that the local properties of the core are dictated by the global
flow properties.
\section{Core Mass Function}\label{sec.massdist}
One of the open questions in star formation is the origin of the
stellar initial mass function (IMF).  \citet{Salpeter55} first
measured this and fit it to a power law,
\begin{eqnarray}
dN &= 0.03 (\frac{M}{\msun})^{\alpha} dM.\\
\alpha &=-2.35
\end{eqnarray}

This fit was done between 1 and 10 $\msun$.  The exact value of the exponent in
the 1 to 10 $M_\odot$ range is still under investigation \citep{Scalo05}, though
recent measurements give the range of $\alpha$ to be between -2.3 and -2.8.

It has been proposed that the IMF and CMF are directly related to one another,
either directly \citep{Motte98} or with some fraction of each core lost in the
final collapse and accretion phase \citep{Enoch08}.  This implies that the IMF
is determined by the global or large scale processes of star formation, in our
model the combined effects of turbulence and gravity, as in the model of
\citet{Padoan02}.  Alternative models have
the IMF set by local physics, once protostars have formed within the prestellar
condensations.  These models include the competitive accretion model of
\citet{Bonnell01}, wherein the population of neighboring protostars influences
the final mass of any given star; and models of \citet{Shu87} or
\citet{Myers10}, where protostellar outflows halt or slow the inflow of gas
onto the protostar.  

\begin{figure} \begin{center}
\includegraphics[width=\hw\textwidth]{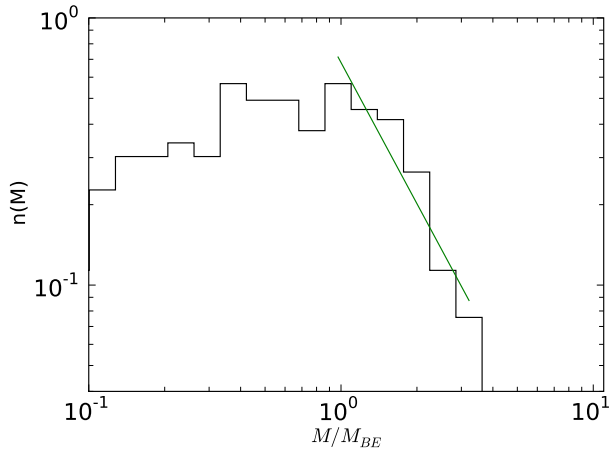}
\caption[ ]{Mass distribution of bound cores.  The fit line is $N(M)\propto M^{-2.1\pm0.6}$, consistent with both IMF and CMF measurements.}
\label{fig.MassDistWithFit} \end{center} \end{figure}

\begin{figure} \begin{center}
\includegraphics[width=\hw\textwidth]{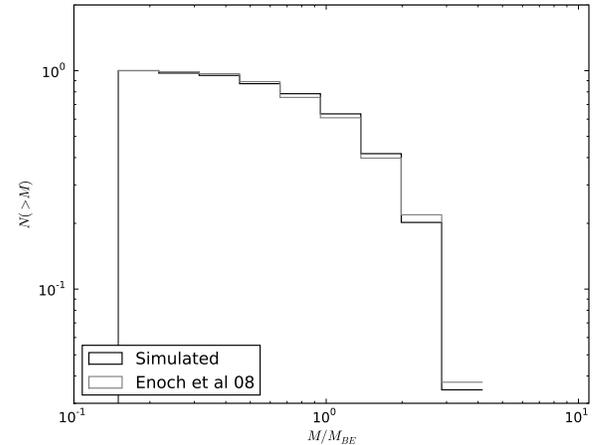}
\caption[ ]{Cumulative mass distributions for our bound cores (black) and the
prestellar cores from
\citet{Enoch08} (grey), relative to the Bonnor-Ebert mass.  Here we have normalized the
observed cores to $M_{BE} = 1.5 \msun$ for comparison}
\label{fig.MassDistCumulative} \end{center} \end{figure}

Figure \ref{fig.MassDistWithFit} shows the mass distribution for all bound
cores.  The fit to the high end of the
distribution is $n(M)\propto M^{-2.1 \pm 0.6}$. The fit was
performed by fitting a power law between the peak and the highest bin for a
succession of bins between 5 and 25.  We find good agreement between our slope
and the IMF slopes mentioned above, and the slope of $-2.3\pm0.6$ measured for
the CMF by
\citet{Enoch08}.
Further agreement with the CMF is seen in figure \ref{fig.MassDistCumulative},
which shows the cumulative mass function $N(>M)=\int_M^\infty n(M) dM$ for our data (black line) and
the prestellar cores from Perseus, Ophiuchus and Serpens presented in
\citet{Enoch09} (grey line).  Here we have
scaled both populations to the Bonnor Ebert mass,
$$
M_{BE} = \frac{1.18 c_s^4}{G^{3/2} \rho_0^{1/2}}.
$$
The observed
points used a Bonnor Ebert mass of $1.5 \msun$, which
corresponds to a background density of $\approx 9000 \rm{cm}^{-3}$ at 10 K.
This is somewhat higher
than the mean density in these clouds, but not unreasonable for the ambient
density immediately surrounding the cores, which have mean densities of
$\approx 10^5 \rm{cm}^{-2}$.  The simulated cores used the mean density in
the box, which in the scaling used throughout this paper gives $M_{BE} = 10
M_\odot$, though strictly speaking this is a free value.  The observational data are all prestellar
cores from the 1.1 mm Bolocam survey of Perseus, Serpens and Ophiuchus 
presented in \citet{Enoch06}, \citet{Young06}, \citet{Enoch07}, respectively,
that do not have an associated infrared
source in the \textit{Spitzer} c2d catalog.  The majority of
these objects can reasonably be assumed to be self gravitating, based on
comparisons of the Perseus cores to kinematic information from the molecular
line survey of the same region by \citet{Rosolowsky08b}.  As our simulation
only attempts to model the prestellar core phase of star formation, and not the
formation of the actual star itself, this sample of objects is the best
observational counterpart for comparison.  


\section{Conclusions}\label{sec.conclusions}
In this work, we present density and magnetic field distributions for a super
Alfv\' enic turbulence simulation with self gravity.   The simulation was run
with the AMR extension of Enzo described by \citet{Collins10}, allowing us
unprecedented spatial resolution.  \Sa\ turbulence
has been proposed as the primary mechanism for star formation, providing good
explanations of the star formation rate \citep{Krumholz05,Padoan10} 
and initial mass function
\citep{Padoan02}.  Here we provide two checks of this model against observations, and
explore deviations from the predictions of \sa\ turbulence caused by the addition of self
gravity.

We find in Sections \ref{sec.densitypdf} and \ref{sec.magneticpdf} that power law
tails develop for both high density and high magnetic field in volume and mass
weighted PDFs, $P(\rho)\propto \rho^{-1.67}$ and $P(B)\propto B^{-2.74}$,
respectively.  The volume-weighted density PDF is consistent with the prediction
of a singular isothermal sphere (SIS) \citep{Kritsuk10} $P(\rho_{\rm{SIS}}) \propto
\rho^{-1.5}$.  

The relationship between the magnetic field and the density also shows a power
law behavior $B\propto \rho^{0.48}$ throughout the gas, consistent with the
findings of \citet{Li04}, who found a similar behavior in the peak density/field
relation in cores.  This then allows us to explain the magnetic PDF, by
combining this result with the density PDF.  

Gravitationally bound cores found in our simulation were compared against several observational
surveys.  Comparisons with the most recent Zeeman splitting measurements of
\citet{Troland08} and \citet{Falgarone08} show that the mass-to-flux ratio in
our simulations agrees in value and behavior with those found observationally.
The relationship between field strength and column density is fit to a power law,
$B\propto N^{0.57}$, demonstrating that significant mass-to-flux, thus magnetic
support, is lost due to motion along the field lines.  

Comparing our core mass function (CMF) to that of prestellar cores in \citet{Enoch08},
we again find excellent agreement.  A slope of $2.1 \pm 0.6$ agrees with their
fit value of $2.3 \pm 0.6$, and cumulative mass distributions line up almost
identically.  The relatively good match between the observed CMF and the
observed IMF indicates that the IMF is determined well before the onset of
nuclear burning, at a relatively low (compared to the protostar) density.  A
multiplicative offset of $>1/4$ is seen between the cores of \citet{Enoch08} and
the observed IMF, indicating that as much as $3/4$ of the mass is lost in the
final collapse phase.  However, this is only an upper limit to the lost
fraction, as the peak of the observed CMF is heavily influenced by its
completeness limit.  Bound cores in our simulation agree with the observed CMF
extremely well, indicating that \sa\ turbulence and gravity are primarily
responsible for the structure of the mass distribution of the CMF, and
ultimately the IMF.

The authors would like to thank A. Kritsuk for his input and many useful
discussions.  The authors would like to acknowledge financial support from NSF grants AST0808184 and
AST0908740, and computational resources provided by the National
Institute for Computational Sciences under LRAC allocation MCA98N020
and TRAC allocation TG-AST090110.


%
%

\bibliographystyle{apj-2009-05-18}
\bibliography{apj-jour,ms}  

\end{document}